\documentclass[a4]{article}
\usepackage{amsmath,amssymb}
\usepackage{graphicx}
\usepackage[hmargin=3cm,vmargin=3.5cm]{geometry}
\begin{document}
\title{Liquid Scintillator Time Projection Chamber Concept}

\author{N. McConkey\footnote{Corresponding author: n.mcconkey@warwick.ac.uk},
  Y.A. Ramachers\\
Department of Physics, University of Warwick, \\Coventry, CV4 7AL, UK}
\date{}
\maketitle
\begin{abstract}
Results are presented from a small-scale experiment to investigate the use of
room temperature organic liquid scintillators as the active medium for a time
projection chamber (TPC). The optical properties of liquid scintillators have
long been known, but their 
ability to transport charge has remained, until now, largely untested.  The
idea of using room temperature liquids as an active medium for an ionisation
chamber was first presented in \cite{EnglerTMS}.  Since then the range of liquid
scintillators available has been greatly developed. We present successful
transport of ionization charges in a selection of both, pure organic liquid
solvents and liquid scintillator
cocktails over 20$\,$mm using a variety of electric drift field strengths. The
target of this research is to offer a cost effective alternative to liquid
noble gas detectors in neutrino physics.
\end{abstract}

\section{Introduction}
A detector capable of delivering the neutrino physics program of the future
will need to be extremely large in target mass, but also have fine grained
tracking across the whole volume.  A liquid argon time projection chamber fits
these criteria well, and has been very successfully demonstrated to work at
the 600$\,$ton scale by ICARUS \cite{ICARUS}. In terms of practicality, however,
there are a number of drawbacks to using 
cryogenic liquids on a scale larger than this, i.e. the 100kton size that is
aimed for in neutrino physics.  Both the cryogenic infrastructure and the
difficulty of purification to the high degree needed prove challenging.

We propose to consider the possibility of a room temperature liquid with
scintillation properties similar to liquid argon, i.e. modern organic liquid
scintillators. 

\subsection{History}
The first observations of the presence of free electrons in a room temperature
liquid were made in 1968, in neopentane \cite{neopentenepaper}, and
tetramethylsilane \cite{oldtmspaper}, but it was not until the early \textquoteleft80s that
these liquids were tested further, in the context of a liquid ionisation chamber
\cite{EnglerTMS}, \cite{HolroydTMS}.  Tetramethylsilane (TMS) and
tetramethyltin (TMT) were considered as potential candidates for the detector
medium.  These chemicals in particular were chosen due to the compact and spherical
shape of the molecules.  Theory suggested that mobility was inversely proportional to
the length of the molecular chain \cite{Adamczewski}, therefore these compact
molecules were likely to have more useful charge transport properties.

\subsection{Organic Liquid Scintillators and Solvents}
In the past 20-30 years there has been a lot of development in organic liquids
for scintillation counting applications There are now many \textquotedblleft safe" solvents
produced, some of which are widely used in particle physics.  Linear alkyl
benzene (LAB) is used in the SNO+ \cite{SNOplus}, Daya bay \cite{dayabay} and
RENO \cite{reno} experiments, phenyl xylyl ethane (PXE) was considered to be
used in Borexino\cite{Borexino}.  Di isopropyl naphthalene (DIN) has not been
used in any particle physics experiments, but is a solvent especially designed
for scintillation counting \cite{DINinvention}, and has optical properties comparable to the other solvents mentioned.

If any one of these liquids is found to be a suitable material for use in a
time projection chamber, the room temperature operation would allow many
practical benefits.  The infrastructure would be much more
straightforward than a system incorporating cryogenics, and there is already
expertise in cleaning organic liquids to a high level of purity
\cite{SNOplus}.

Whilst a large volume detector is necessary for long baseline neutrino
experiments, a liquid scintillator TPC could also have a part to play in the
search for neutrinoless double beta decay.  Loading with
isotopes is, again, well researched in organic liquid scintillators.  With a
large volume loaded liquid scintillator TPC, it would be possible to combine
both high masses and fine grain tracking for double beta decay searches, a
merger of the SNO+ \cite{SNOplus} and SuperNEMO \cite{snemo} concepts.  

\subsection{Feasibility}
There are many measurements which need to be made in order to determine
whether organic liquid scintillators represent a viable alternative to
cryogenic liquids for use in TPCs, but the initial task is to investigate
whether or not they can transport charge in any practically relevant manner.

For this purpose a gridded ionisation chamber was constructed, with an initial
drift distance of 22$\,$mm.  An Am$\,$241 alpha source is mounted at the cathode, and
a Frisch grid, mounted 7$\,$mm from the readout anode is resistively coupled to
the anode, keeping the ratio of the two electric fields constant at 53:50,
which makes the grid optically transparent to drifting electrons
\cite{oldargonpaper}. The 
measurements presented here are the first results from the measurement 
programme using this detector.

\section{Selected Results}

\begin{table*}[ht]
\hfill{}
\begin{tabular}{|l|c|c|}
\hline
Chemical name & Chemical Formula & Molecule structure \\
\hline
Di isopropyl naphthalene (DIN) & $C_{16}H_{20}$
& \includegraphics[height=0.6cm]{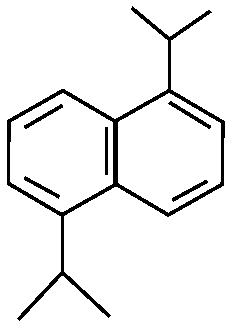}\\ 
\hline
Mono isopropyl naphthalene (MIPN) & $C_{16}H_{14}$
&\includegraphics[height=0.6cm]{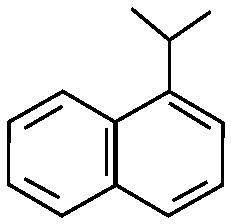} \\ 
\hline
Mono isopropyl biphenyl (MIBP) & $C_{15}H_{16}$
&\includegraphics[height=0.6cm]{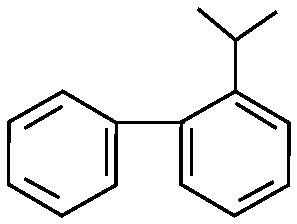} \\ 
\hline
Linear alkyl benzene (LAB) & $C_6H_5C_{10}H_{21}$ to $C_{13}H_{27}$
&\includegraphics[height=0.6cm]{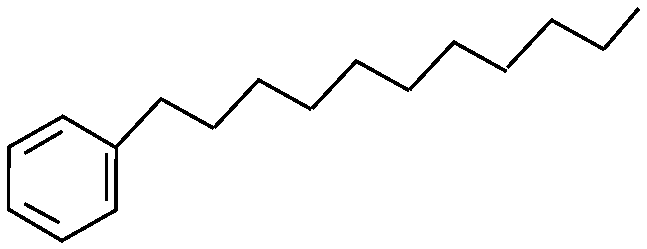} \\ 
\hline
Phenyl xylyl ethane (PXE) & $C_{16}H_{18}$
&\includegraphics[height=0.6cm]{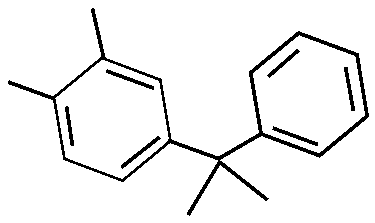} \\ 
\hline
\end{tabular}
\hfill{}
\caption{The organic liquids and solvents tested and their molecular structure.}
\label{tablelist}
\end{table*}

\subsection{Di isopropyl naphthalene (DIN)}

Two types of di isopropyl naphthalene have been tested, both the pure solvent,
and an off-the-shelf scintillation cocktail, i.e. DIN with wavelength shifting
fluors added.

The event rate measurements from the two liquids are strikingly different; the
pure liquid has about twice the number of events than measured in the
scintillation cocktail, for any given drift field.  This suggests that the
presence of the fluors in the liquid has a detrimental effect on the charge
transport properties of the liquid.  

\begin{figure}[h]
\centering
\includegraphics[width=9cm]{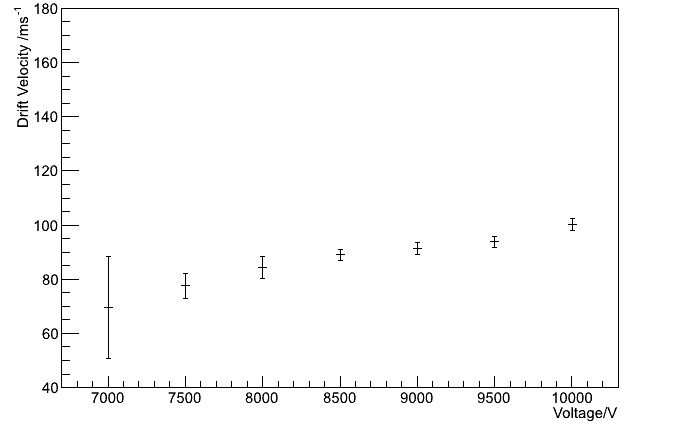}
\caption{Drift speed as a function of voltage, in di isopropyl naphthalene
  (DIN).} 
\label{dinvelocity}
\end{figure}

The drift speed in DIN was measured to be $(101\pm2)\,$ms$^{-1}$ in a
drift field of 6.93$\,$kVcm$^{-1}$, as shown in Figure \ref{dinvelocity}.
This is approximately a factor of 20 slower than the corresponding drift
speed in liquid argon \cite{aprilebook}, see discussion below.   

\subsection{Mono isopropyl naphthalene (MIPN)}

\begin{figure}[h]
\centering
\includegraphics[width=9cm]{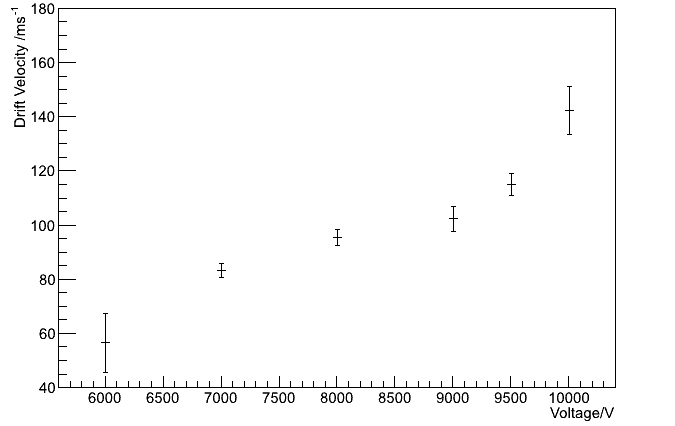}
\caption{Drift speed as a function of voltage, in mono isopropyl naphthalene
  (MIPN).} 
\label{mipnvelocity}
\end{figure}

Mono isopropyl naphthalene is an organic solvent with similar chemical
structures and properties to DIN, although it is not a commonly used liquid
for scintillation counting. 
The drift speed of MIPN as a function of drift voltage is shown in Figure
\ref{mipnvelocity}.  At an electric field of 6.93$\,$kVcm$^{-1}$, the drift
speed has been measured to be $(143\pm9)\,$ms$^{-1}$. 

\subsection{Mono isopropyl biphenyl (MIBP)}

Mono isopropyl biphenyl is an organic solvent which is also not used for
scintillation counting, therefore its optical properties are currently
unknown, but its chemical similarity to DIN warranted testing.

\begin{figure}[h]
\centering
\includegraphics[width=9cm]{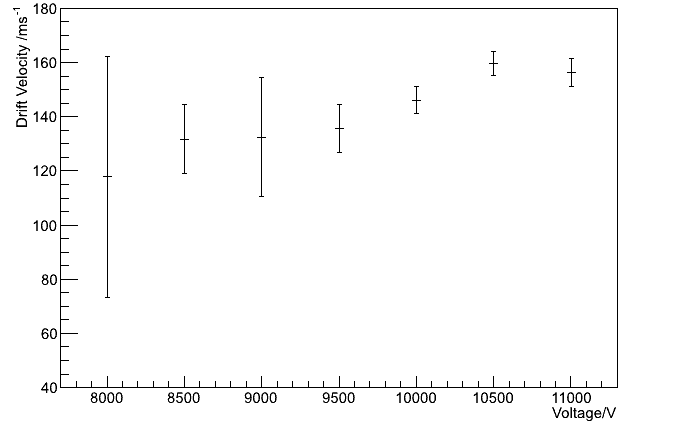}
\caption{Drift speed as a function of voltage, in mono isopropyl biphenyl
  (MIBP).} 
\label{mibpvelocity}
\end{figure}

The drift speed of MIBP as a function of drift voltage is shown in Figure
\ref{mibpvelocity}. At a drift field of 7.63$\,$kVcm$^{-1}$, the drift speed
has been measured to be $(156\pm6)\,$ms$^{-1}$. 

\subsection{Linear alkyl benzene (LAB)}

The solvent linear alkyl benzene has been obtained with scintillation fluors
pre-mixed.   
It has been found that although charge transport is possible in LAB, the
dielectric strength of the material is much lower than the other solvents
tested, therefore it is not possible to reach the electric fields needed in
order to drift charge at a measurable speed.   

\begin{figure}[h]
\centering
\includegraphics[width=9cm]{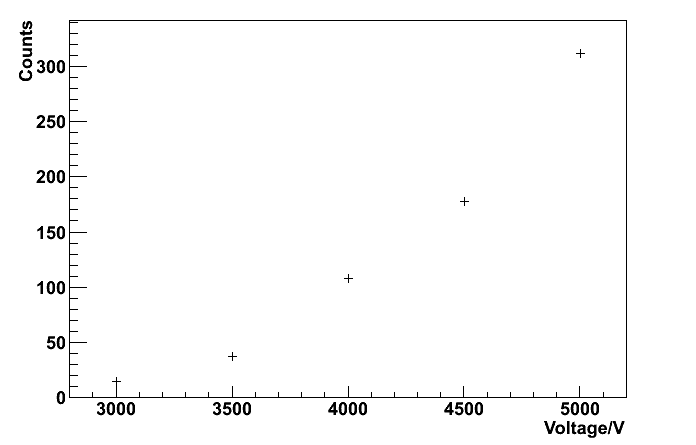}
\caption{The event rate in linear alkyl benzene as a function of drift
  voltage. Note the relatively low applied maximum bias before onset of
  sparking in this liquid.}
\label{labrate}
\end{figure}

It is also suggested \cite{Adamczewski} that the molecular shape (see Table
\ref{tablelist}) of LAB would not correspond to one which would have a high
mobility, and therefore the low event rates recorded (see Figure
\ref{labrate}) are due to the nature of the solvent as well as the fluors
present. 

\subsection{Phenyl xylyl ethane (PXE)} 
\begin{figure}[h]
\centering
\includegraphics[width=9cm]{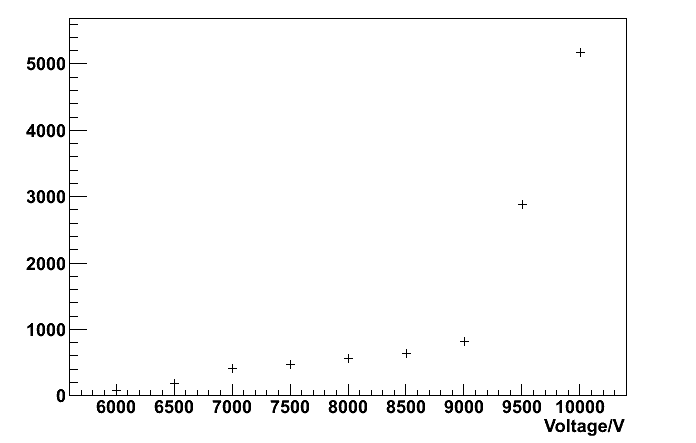}
\caption{The event rate in phenyl xylyl ethane as a function of drift voltage.}
\label{pxerate}
\end{figure}

The solvent phenyl xylyl ethane has been obtained with scintillation fluors
premixed.  
The liquid has been found to break down at voltages above 10kV (see Figure \ref{pxerate}).  The plateau between 7 and 8$\,$kV, with a low event rate seems promising; it is possible that as in DIN, the fluors are reducing the event rate.  It is intended to source this PXE as a pure solvent without fluors for further testing. 

\section{Conclusions}
With the exception of LAB, all tested solvents and cocktails are able to
transport charge.  This discovery gives a clear way forward; a measurement
program to fully characterise their transport properties. 

A drift speed of roughly a factor 20 less than the typical drift speed
observed in liquid argon detectors at around 0.5 to 1$\,$kV/cm drift field
strength does not invalidate detector operation for room-temperature
liquids. The only direct consequence would be a constraint on permissible
total event rates which need to be kept low generally for rare event searches
such as neutrino detection anyway. However, low drift speed suggests low mean
free paths for charge transport for the simple reason that charges should
have a higher probability to be trapped by impurities the longer they travel
in the liquid. This mean free drift length will be addressed in
our measurement programme in the near future.

The preliminary drift velocities will be refined, and the mean free path
measured, as the drift distance in the current setup is increased up to
100$\,$mm. It is planned to implement a pulsed photoconversion electron
source. With the advent of cheap and fast UV LED light sources \cite{source},
such a source is expected to permit precision control of charge--pulse
production in our setup. It
is also necessary to characterise fully the scintillation  
properties for each solvent as opposed to commercial cocktails, using for
instance optical spectroscopy in the UV spectral range.

This further work will determine whether or not the organic liquids presented
will prove a viable alternative to cryogenic liquids for use in large volume
TPCs with far reaching consequences for neutrino physics.

\end{document}